\documentclass[a4paper,12pt]{article}
\usepackage[utf8]{inputenc}
\usepackage{lmodern}
\usepackage[T1]{fontenc}
\usepackage{comment}
\usepackage{lscape}
\usepackage{amsmath}	
\usepackage{amssymb}
\usepackage{amsthm}
\usepackage{cases}
\usepackage{graphicx}

\usepackage{caption}
\usepackage[table]{xcolor}

\usepackage[normalem]{ ulem }
\usepackage{soul}

\usepackage{multirow}
\usepackage{amsfonts}
\usepackage{float}
\usepackage{diagbox}

\usepackage{fullpage}
\usepackage{url}
\usepackage{rotating}
\usepackage{eurosym}
\usepackage{wrapfig}
\usepackage[final]{pdfpages} 
\usepackage{epstopdf} 
\usepackage{subfig}
\usepackage[a4paper]{geometry}
\usepackage{subfiles}
\usepackage[nottoc]{tocbibind} 
\usepackage{placeins}
\usepackage{float}

\usepackage{listings}
\usepackage{color, colortbl}

\usepackage{hyperref}
\usepackage{xcolor}
\usepackage{titlesec}
\providecommand{\keywords}[1]{\textbf{\textit{Keywords---}} #1}
\usepackage{pdfpages}
\titleformat{\paragraph}
{\normalfont\normalsize\bfseries}{\theparagraph}{1em}{}
\titlespacing*{\paragraph}
{0pt}{3.25ex plus 1ex minus .2ex}{1.5ex plus .2ex}

\hypersetup{
colorlinks,%
citecolor=black,%
filecolor=black,%
linkcolor=black,%
urlcolor=black
}

\newcommand{\argmin}{\operatornamewithlimits{argmin}}

\begin{document}
\title{Real-time Updating of Dynamic Social Networks for COVID-19 Vaccination Strategies}

\author{Sibo Cheng$^{1}$, Christopher C. Pain$^{2}$,
 Yi-Ke Guo$^{1}$, Rossella Arcucci$^{1}$\\
\small $^{1}$ Data Science Instituite, Department of Computing, Imperial College London, UK\\
        \small $^{2}$ Department of Earth Science \& Engineering, Imperial College London, UK\\
}

\date{}

\maketitle

\begin{abstract}
Vaccination strategy is crucial in fighting the COVID-19 pandemic. Since the supply is still limited in many countries, contact network-based interventions can be most powerful to set an efficient strategy by identifying high-risk individuals or communities. However, due to the high dimension, only partial and noisy network information can be available in practice, especially for dynamic systems where contact networks are highly time-variant. Furthermore,
the numerous mutations of SARS-CoV-2 have a significant impact on the infectious probability, requiring real-time network updating algorithms. In this study, we propose a sequential network updating approach based on data assimilation techniques to combine different sources of temporal information. We then prioritise the individuals with high-degree or high-centrality, obtained from assimilated networks, for vaccination. The assimilation-based approach is compared with the standard method (based on partially observed networks) and a random selection strategy in terms of vaccination effectiveness in a SIR model. The numerical comparison is first carried out using real-world face-to-face dynamic networks collected in a high school, followed by sequential multi-layer networks generated relying on the Barabasi-Albert model emulating large-scale social networks with several communities.
\end{abstract}


\section{Introduction}

The world is still in the midst of the COVID-19 pandemic. 
The World Health Organization (WHO) and partners are working together on the response, tracking the pandemic, providing recommendations on critical steps, delivering necessary medical supplies to those in need and, finally, racing for the development and introduction of safe and reliable vaccines.
Every year, vaccines save millions of lives. Vaccines work to identify and fend off the viruses and bacteria they attack by training and preparing the body's natural defences, the immune system. If the body is eventually exposed to such disease-causing germs, it is ready to kill them instantly, avoiding illness.
By the end of July 2021, nearly 300 vaccine candidates for COVID-19 are currently in trials\footnote{\url{https://www.who.int/publications/m/item/draft-landscape-of-covid-19-candidate-vaccines}}, and several of them,  such as AstraZeneca, Pfizer, Moderna and Gamaleya, have already been distributed in all countries to protect individuals. 
No other vaccine in human history has been so eagerly anticipated, especially given that until now no drugs are demonstrated to be available to treat COVID-19. 
By July 30th 2021, almost or above $50\%$ of the population has been fully vaccinated in North America and European countries, including the USA($50.2\%$), the UK($57.2\%$) and Canada($59.6\%$). However in some less developed nations the vaccination rate is worryingly low such as India($7.6\%$) and Peru($14.7\%$), both having experienced a major COVID crisis recently.
Since the vaccination capacity in these countries remains limited until now, people who are most at risk, such as healthcare workers and older population \cite{Mills2021}, are given priority \cite{kumar2021strategy}. The effectiveness of the current vaccinations in addressing newly developed virus variants (e.g.,B.1.617.2 (Delta) and C.37 (Lambda)) has also been challenged~\cite{lopez2021effectiveness}, leading to the possibility of requiring new vaccinations or doses\footnote{\url{https://www.gov.uk/government/publications/long-term-evolution-of-sars-cov-2-26-july-2021}}.
Vaccination strategies play an essential role in preventing the rapid diffusion of COVID-19. Clustering analysis has investigated transmission cascades in local social communities.  Among all connecting clusters, particular attention has been given to educational settings, including high schools and universities \cite{Ismail2020}. Much effort has been devoted to maintaining the possibility of face-to-face teaching during the pandemic. However thousands of clusters and outbreaks of COVID-19 have been reported in educational establishments. As mentioned in \cite{kumar2021strategy}, the Delta variant has become the dominant strain in the UK, spreading rapidly in schools since May 2021. Hence, finding an optimal vaccination strategy for students and staff has become vital to protecting children and young people since many countries, including India and the UK, plan to reopen colleges and schools, either in full or in part, from September 2021.\\

Continuous effort has been made for several decades to develop the simulation of infectious diseases based on observed social networks\cite{camacho2020four}, including, for instance, H1N1 influenza (face-to-face contact network)~\cite{Cauchemez2011} and HIV (sexual contact network)~\cite{Keeling2005}. Social network-based analysis for disease spread modelling has been widely implemented since the outbreak of COVID-19 \cite{Mauras2020, Firth2020}, with the help of SIR (Susceptible-Infected-Recovered) or SEIR (Susceptible-Exposed-Infected-Recovered) models.
 When the network structure of contacts is (at least) partially observable, network-based interventions are most helpful in determining an optimal vaccination strategy under a limited capacity, which has been proved in a variety of infectious diseases~\cite{Meyers2006}, \cite{Rushmore2014}. These strategies are usually based on some individual-level measures, such as node degree or graph centrality, which require knowledge of the full network. Furthermore, significant variance of COVID infection probability is also observed~\cite{Davies2020} according to ages and activities. Meanwhile, many connecting clusters of COVID-19 have been identified in schools and workplaces\cite{Yong2020}, where individuals share similar characteristics. Thus the infectious probability of intra-connections inside these clusters could be considered homogeneous. This fact leads to the idea of multi-layer network modelling where the infectious probability may vary from layer to layer.\\
 
 Much effort has been given to using network-based information for formulating optimal policy responses to COVID-19 \cite{LALMUANAWMA2020,sym12101646}, including social distancing and countrywide lockdown \cite{Block2020}. However, the observation of social networks is often noisy (with either missing connections or mistaken edge weights), and, most of the time, incomplete \cite{Rushmore2014}. Obtaining precise knowledge is particularly challenging since face-to-face contact networks are strongly time-variant. The noise-level could be up to $74\%$ (missing edges) for observed connection networks, as mentioned by \cite{KOSKINEN2013514}. On the other hand, as pointed out by \cite{alsdurf2020covi}, contact tracing applications can significantly reduce the rate of infection in the studied population when the participation rate is above $60\%$. In other words, it is critical to maintaining an error level inferior to 40\%. Therefore, a considerable gap can be found between the required precision and the available data on the temporal networks. real-time updatings of prior network knowledge is thus essential to improving vaccine efficiency.\\
 
 In this paper, by investigating how the accuracy of network data could impact vaccination effectiveness, we propose a real-time network updating approach based on sequential data assimilation (DA) techniques. Originally developed in the field of meteorological and environmental science, DA has been applied to a wide variety of industrial domains, including geophysical modelling~\cite{Carrassi2017}, hydrology~\cite{cheng2020b} and economics~\cite{Nadler2019}. Recently, sequential DA algorithms have also been used for real-time parameter identification in the SIR model for COVID spread simulation~\cite{Wang2020,Nadler2020,Evensen2020}. An important advantage of using DA, compared to other statistical models for network reconstruction(e.g \cite{Peixoto2019}) is that DA is widely used for large-dimension problems with noisy and limited prior data. As an example, Graph Neural Networks (GNN) \cite{wu2020comprehensive} have been demonstrated to have high accuracy in network reconstructions with missing data \cite{you2020handling}. However, this approach requires retraining for each temporal graph, leading to difficulties in real-time predictions.
DA and dynamic network data have been combined in \cite{cheng2020graph} where the authors propose a graph clustering approach for the efficient localization of error covariances within an ensemble-variational DA framework. \cite{ihler2007graphical} presents a relationship between
statistical inference using graphical models and optimal sequential estimation algorithms such as Kalman filtering. 
In this work, DA is employed for real-time updating of the network, including novel information from dynamic observations. This contributes to leveraging the information embedded in different noisy/incomplete observations using an optimisation process to reconstruct the current network. This is computationally feasible for large-scale problems thanks to the sparsity of the contact networks. Here, we propose two DA models for different parametrizations: 
\begin{enumerate}
    \item The first consists of reconstructing the complete contact network structures by observing the edges in temporal sub-networks (as described in Sect.~\ref{sec: school});
    \item The second adjusts inhomogeneous infectious probabilities in a multi-layer network modelling (as described in Sect.~\ref{sec: multilayer}).
\end{enumerate}
These two models are respectively applied to 
\begin{enumerate}
    \item A real-world dynamic network dataset describing the contacts of French high school students in a week \cite{Genois2018}, collected using wearable sensors;
    \item Generated scale-free multi-layer networks, where each layer represents a social community/cluster, determined by individual characteristics such as age or activity.
\end{enumerate}

Preliminary analysis is performed to understand the data structure (clustering, classes, grades) of the high school contact networks and to demonstrate the time-variance. The same data set, collected in a high school in Lyon, has been used to simulate a COVID outbreak and estimate the reproductive ratio $R_0$ in \cite{Mauras2020}. It is also shown in their work that the study of contact networks in schools or workplaces could lead to more optimal contact-limiting strategies, such as self-isolation or countrywide lockdown. In this work, we make similar assumptions to \cite{Mauras2020} in terms of infection rate (slightly higher regarding new SARS-CoV-2 variants) in the contact network. However, since the availability of the temporal network data is limited, we set a small value for the average recovery period (5 days) to simulate the highest number of infected in the SIR model. With regard to multi-layer systems, the dynamic networks are generated using the Barabasi-Albert model \cite{Albert2002}, with a power law degree distribution. The latter exists widely in real social networks. Since mutations of SARS-CoV-2 have continuously arisen, the infection probability in each network layer is supposed to be time-variant, following an additive stochastic process.
In both cases, the SIR simulation is carried out with realistic assumptions of COVID-19 to simulate the SARS-CoV-2 propagation, while real-time observations are generated synthetically based on preliminary network analysis. The DA models proposed in this paper are general, and could be applied to various scenarios with different types of real-world dynamic networks and observation data. \\

In summary, in this work we
\begin{itemize}
    \item simulate the COVID-19 propagation and vaccination impact using real or generated multi-layer networks with the SIR model. 
    \item propose a DA framework, with two different network parametrizations, to sequentially update the network structure based on noisy prior information and real-time observations. 
    \item compare different graph measures, such as node degree and betweenness centrality for vaccination prioritization criteria of prior and assimilated networks.  
\end{itemize}

The paper is organized as follows. Sect.~\ref{sec:G_model} introduces the graph-based diffusion modelling and vaccination strategies. Data assimilation principle and adaptation of graph data are presented in Sect.~\ref{sec:da}. Sect.~\ref{sec: school} shows numerical experiments in real-world social contact networks, and Sect.~\ref{sec: multilayer} shows experiments with multi-layer networks. Sect.~\ref{sec:conclusion} closes the paper with conclusions and future work.

\section{Graph-based diffusion modelling and vaccination strategies}
\label{sec:G_model}
\subsection{SIR model}

\begin{figure}[H]
  \centering
    \includegraphics[width = 5. in]{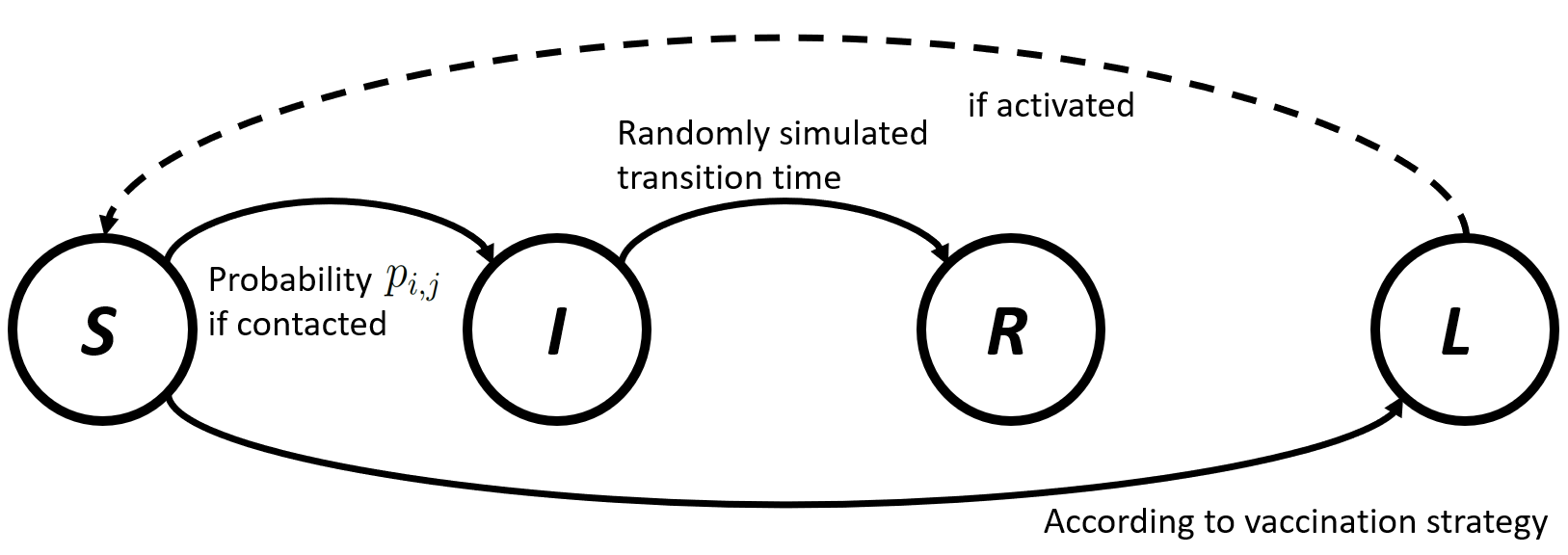}
    \caption{Illustration of network-based SIR model with a vaccination state $L$.}
  \label{fig:network_SIR}
\end{figure}

The analysis of the diffusion is conducted using a standard SIR model \cite{anderson1991discussion} with an additional state describing the number of vaccinated people, as shown in Fig.~\ref{fig:network_SIR}. For each individual, $S,I,R$ denote the susceptible, the infected and the recovered (patients who are not infectious anymore). The SIR assumption has been widely adapted to simulate COVID-19 propagation \cite{COOPER2020,Wang2020,venkatasen2020forecasting} since reported COVID reinfection cases (e.g \cite{Tillett2021}) are still rare compared to the total number of reported cases thus far. The SIR model has also been broadly used in network-based disease simulations via random-walk-based simulations \cite{Keeling2005}. Each node symbolizes an individual in the social network, whose status can alter from susceptible to infected (S-I), or infected to recovered (I-R), according to the random walk through temporal edges \cite{Durrett4491}. The transition from susceptible (S) to vaccinated (L) only takes place when required according to chosen vaccination strategies. In contrast to classical disease modelling, since recent research\cite{lopez2021effectiveness} shows that current COVID vaccinations can be significantly less effective when facing new variants (e.g.,B. 1.617. 2 (Delta)), the L-S and L-I transitions can be activated as shown in Fig.~\ref{fig:network_SIR}. More details about the transition probabilities are given in Sect.~\ref{sec: Graph-based vaccination}. In view of the fact that until these days the infection probability after vaccination is still unclear, L-S and L-I transitions are not considered in this study. Nevertheless, the developed model can easily incorporate these types of transitions when required. 

\subsection{Graph-based vaccination strategy}
\label{sec: Graph-based vaccination}
Both disease spread simulation and optimal vaccination modelling based on social networks have been receiving increasing interest for different types of infectious diseases \cite{Newman2002}. We  consider an undirected graph $\mathcal{G}$ that is a pair of sets $\mathcal{G}= (V,E)$, where $V = \{v_1, v_2 ... v_n\}$ represents the set of individuals (graph nodes) and the set $E$ contains the edges, each connecting a pair of individuals. Each graph edge $e \in E$ is represented by a triple $e = (v_i,v_j, w_{i,j})$ where $v_i,v_j $ are the two endpoints and $w_{i,j} \in \mathbb{R}$ is the edge weight. For unweighted graphs  $w_{i,j} \in \{0,1\}$, while for weighted graphs $w_{i,j}$ could represent the frequency or the intimacy of the contact. In epidemic spread modelling, the infectious probability $p_{i,j}$ from the individual $i$ to $j$ (and vice versa) is often in function of $w_{i,j}$,  $p_{i,j} = \mathcal{IP}(w_{i,j})$. We also note that $p_{i,j}$ may depend on individual-level characteristics of $v_i$ and $v_j$, such as age or activities. The connecting graph can be fully represented by the associated adjacency matrix $\mathbf{A}= \{ A_{i,j}\}_{i,j=1,\dots,n}$. We use three Boolean vectors $\{\mathbf{I}_t, \mathbf{L}_t, \mathbf{R}_t\} \in \{ \{0,1\}^n\}^3$ to indicate the status of each individual, either infected, vaccinated or recovered in the SIR model, at time $t$. The recovery period $T_{\gamma} \in \mathbb{N} $ is an uniform distributed random variable generated individually for each individual.\\

If we adopt the edge-wise function $\mathcal{IP}(.)$ in the whole network,

\begin{align}
    \mathcal{IP}(\mathcal{G})_{i,j} = \mathcal{IP}(A_{i,j}),
\end{align}

the infectious probability vector $\mathbf{I}^p_t \in (0,1)^n$ at time $t$ in this SIR model reads

\begin{align}
    \mathbf{I}^p_t = \big( \mathcal{IP}(\mathbf{A}_{t-1}) \hspace{1mm} \mathbf{I}_{t-1} \big) \odot (\mathbf{1}_n - \mathbf{R}_{t}) \odot (\mathbf{1}_n - \mathbf{L}_{t-1}) \odot (\mathbf{1}_n - \mathbf{I}_{t-1}), \label{eq:SIR_graph}
\end{align}

where $\mathbf{1}_n = [ 1,1 ...1]^T$  and $\odot$ denotes the vector-wise Hadamard product. \\

Following a uniform probability distribution, the vector of infections $\mathbf{I}_t$ is simulated using $\mathbf{I}^p_t$ and $\mathbf{I}_{t-1}$. The only controllable variable in Eq.~\ref{eq:SIR_graph} is the vaccination vector $\mathbf{L}_t$.\\

Different graph-based vaccination strategies can be employed to enhance the immunization impact with a limited vaccination capacity. The state of the art approaches are usually determined by observed individual- or community- level social connections, often involving classical graph measures, for instance, graph degree, betweenness centrality \cite{Freeman1997} or community links\cite{Yiping2008}. Much efforts have also been made to use these strategies in practical settings where significant positive impacts have been observed \cite{harling_onnela_2018}. Since the available graph data often include non-negligible uncertainties (missing vertices or edges), statistical models are commonly employed to provide an optimal estimation of these graph measures.  Practical approaches involve, for example, fixed choice designs (FCD) \cite{MCCARTY2007300} and the nomination strategy \cite{Gracia2017}, both based on an estimation of the graph degree. Even with partially observed dynamic networks, the vaccination strategy could be significantly improved in terms of reducing the maximum infected number and delaying the disease propagation, compared to a random choice \cite{YANG2019115}. Nevertheless, precise knowledge of the network structure is crucial to determining an efficient vaccination strategy. It is essential to use community-based approaches (e.g \cite{Genois2014}, \cite{Yiping2008}), since graph clustering algorithms can be sensitive to noises. However, the data collection of dynamic social networks remains cumbersome, especially for large dimensional problems.  In this paper, we conducted our analysis based on three classical strategies, considered less sensitive to data noise, compared to community-based approaches.

\subsubsection*{Random}
The individuals to be vaccinated are randomly chosen according to the number of doses limited, where no network knowledge is used. 
\subsubsection*{Highest degree}
For each temporal network, we choose to vaccinate people with the most contacts based on prior knowledge. Only observable individuals are taken into account. The degree $d(v)$ of node $v$ in a network is simply defined as the sum of the column (or the row for undirected graphs) of the adjacency matrix,

\begin{align}
    d(v) = \sum_{k = 1}^n |A_{k,v}|
\end{align}
\subsubsection*{Highest Centrality}
The betweenness centrality \cite{Freeman1997} $g(v)$ of node $v$ is defined as the number of shortest paths of all pairs of nodes in the graph that pass by the node $v$,\\
\begin{equation}
    g(v)=\sum\limits_{u\neq q \neq v}\frac{\sigma^\mathbf{A}_{uq}(v)}{\sigma^\mathbf{A}_{uq}} \quad u,q \in V
\end{equation}
 where $\sigma^\mathbf{A}_{uq}$ represents  the total number of shortest paths from node $ u$ to node $ q $ and $ \sigma^\mathbf{A} _{uq}(v)$ is the number of those paths that pass through $ v$.\\

Other graph measures relying on detailed understandings of the network (e.g \cite{Yiping2008}) could also be used to establish a vaccine strategy. However, in real applications precise knowledge of the network is often out of reach. Here, our criteria for choosing graph-based vaccination strategies are two-folds: computationally efficient and non-sensitive to observation noise. The latter ensures the "validity" of the methodology even when working with incomplete networks. To enhance our estimation of dynamic contact networks, we make use of data assimilation algorithms.

\section{Data assimilation principle and adaptation of graph data}\label{sec:da}

In this section we introduce the variational data assimilation concept and the resolution using a linear estimator. We also introduce the novel approach which combines DA techniques with dynamic network data.

\subsection{Variational assimilation and BLUE}
DA algorithms aim to combine different sources of noisy information in order to provide a more reliable estimation of the current system. The state variables could be either a physical field or a sequence of parameters. The true state, denoted by $\textbf{x}^\textrm{true}$, stands for the theoretical value of the state at some given coordinates/time, often out of reach in real-world applications. The objective of the assimilation is to gain an optimal approximation $\textbf{x}^a$ of the true state $\textbf{x}^\textrm{true}$, based on the prior information which are two parts: an initial state estimation $\textbf{x}^b$ (so-called the background state)  and  an observation vector $\textbf{y}$.
The former is often issued from prior numerical simulations/predictions while the latter can be obtained via physical measures of some control variables. 
Their tolerances, regarding theoretical values, are quantified by $\epsilon_b$ and $\epsilon_y $, 
\begin{align}
    \epsilon_b &= \textbf{x}^b-\textbf{x}^\textrm{true} \sim \mathcal{N}(0, \textbf{B})\\
    \epsilon_y &=\textbf{y}-\mathcal{H}(\textbf{x}^\textrm{true}) \sim \mathcal{N}(0, \textbf{O}), \notag
\end{align}

where the observation operator $\mathcal{H}$ from the state space to the observable space is supposed to be known. The probability distributions of the prior error are supposed to be centred Gaussian, characterized respectively by the covariance matrices $\textbf{B}$ and $\textbf{O}$.

The key idea in variational methods is to find a balance between the background and the observations using maximum a posteriori (MAP) method \cite{cheng2019}.  This leads to the loss function weighted by the inverse of $\textbf{B}$ and $\textbf{O}$,
\begin{align}
    J_{\text{3D-VAR}}(\textbf{x})&= \frac{1}{2}(\textbf{x}-\textbf{x}^b)^T \textbf{B}^{-1}(\textbf{x}-\textbf{x}^b) + \frac{1}{2}(\textbf{y}-\mathcal{H}(\textbf{x}))^T \textbf{O}^{-1} (\textbf{y}-\mathcal{H}(\textbf{x})) \label{eq_3dvar}\\
   &=\frac{1}{2}||\textbf{x}-\textbf{x}^b||^2_{\textbf{B}^{-1}}+\frac{1}{2}||\textbf{y}-\mathcal{H}(\textbf{x})||^2_{\textbf{O}^{-1}} \label{eq:op_3dvar}.
\end{align}
The optimisation problem defined by the objective function of Eq.~(\ref{eq:op_3dvar}) is called three-dimensional variational method (\textit{3D-VAR}), which can also be considered as the general equation of variational methods without considering the transition model error. The output of Eq.~\ref{eq:op_3dvar} is denoted as $\textbf{x}^a$, i.e.
 \begin{align}
    \textbf{x}^a = \underset{\textbf{x}}{\argmin} \Big(J(\textbf{x})\Big). \label{eq:argmin}
 \end{align}
 
 If $\mathcal{H}$ can be approximated by some linear operator $\textbf{H}$,  Eq.~\ref{eq:argmin} can be solved via BLUE (Best Linearized Unbiased Estimator) formulation,
   \begin{align}
    \textbf{x}^a &= \textbf{x}^b+\textbf{K}(\textbf{y}-\textbf{H} \textbf{x}^b) \\
    \textbf{P}_\textrm{A} &= (\textbf{I}-\textbf{K}\textbf{H})\textbf{B} \label{eq:BLUE}
 \end{align}
 where $\textbf{P}_\textrm{A} = \textrm{Cov}(\textbf{x}^a-\textbf{x}_\textrm{true})$ is the analyzed error covariance and the $\textbf{K}$ matrix, given by
 \begin{equation}\label{eq:Kgain_BLUE}
	\textbf{K}=\textbf{B} \textbf{H}^T (\textbf{H} \textbf{B} \textbf{H}^T+\textbf{O})^{-1}
\end{equation}
is so called the Kalman gain matrix.
In the rest of this paper, we denote $\textbf{H}$ as the linearized transformation operator. The case when $\mathcal{H}$ is non-linear is more challenging for finding the minimum of Eq.~\ref{eq_3dvar}, especially for high-dimensional problems. The resolution often involves gradient descent algorithms (such as "L-BFGS-B" or adjoint-based numerical techniques). \\

\subsection{Online assimilation with graph data}

The essential idea is to perform real-time updating of the partially observed dynamic networks based on other available information, such as sub-graph structures or the current number of those infected. To this end, the prior observed network $\mathbf{A}^b_t$ at time $t$ is considered as the background state (i.e., $\mathbf{x}^b_t = \mathbf{A}^b_t$), while other information is embedded in the observation vector $\mathbf{y}_t$. 

\begin{figure}[H]
  \centering
    \includegraphics[width = 5.2 in]{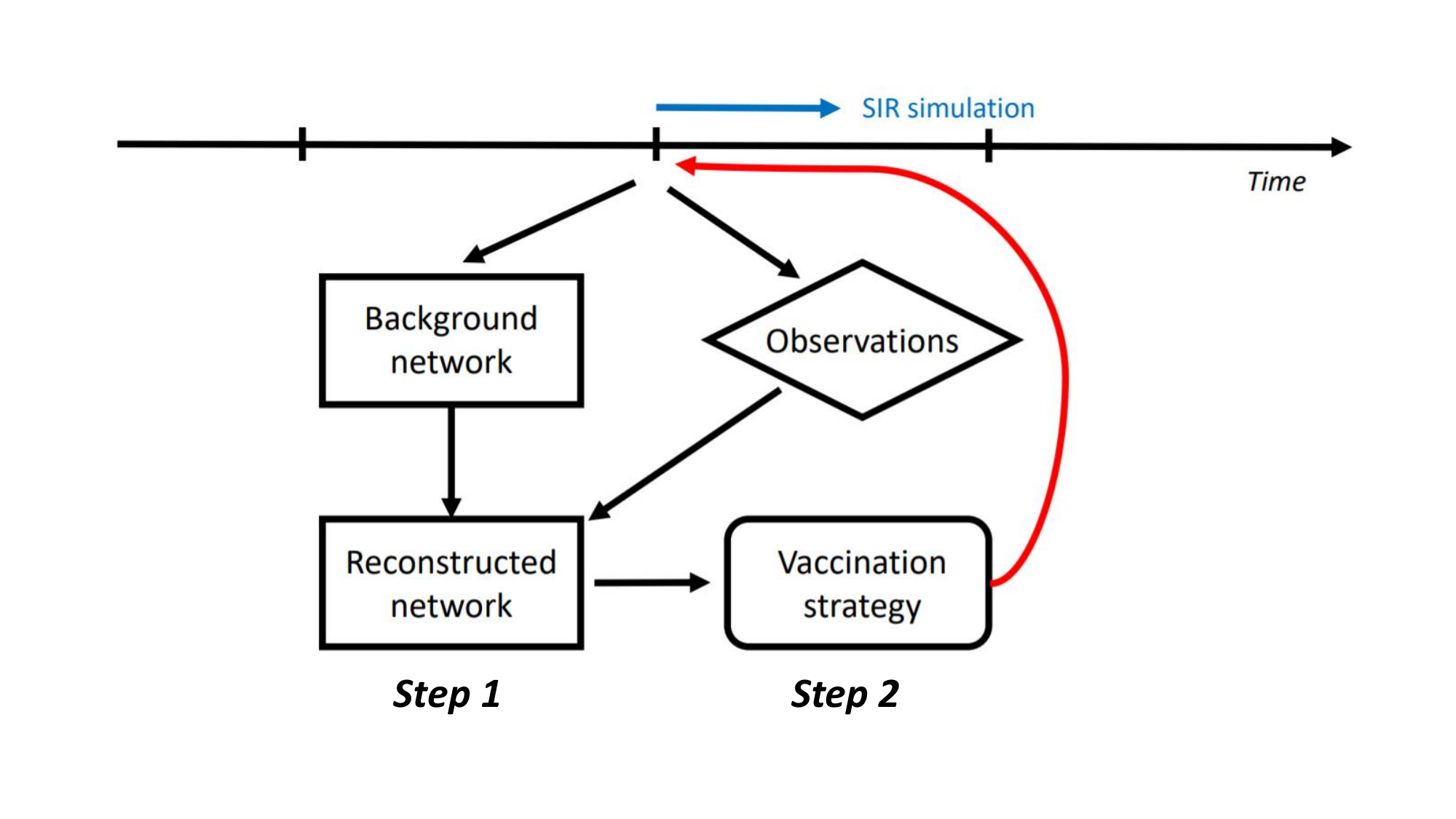}
    \caption{Illustration of real-time DA updating for partially observed contact networks}
  \label{fig:DA_network}
\end{figure}

Once the current contact network is updated based on Eq.~\ref{eq:op_3dvar}, vaccination strategies can be implemented on the analyzed network $\mathbf{x}^a_t = \mathbf{A}^a_t$ (i.e.,step 1 $\rightarrow$ step 2 in Fig.~\ref{fig:DA_network}) which is a more accurate approximation of the true state. The degree and the betweenness centrality of the assimilated network is given by 

\begin{align}
    d^a_t(v) = \sum_{k = 1}^n |{(\mathbf{A}}^a_t)_{k,v}|, \quad     g^a(v)=\sum\limits_{u\neq q \neq v}\frac{\sigma^{\mathbf{A}^a}_{uq}(v)}{\sigma^{\mathbf{A}^a}_{uq}}.
\end{align}

where $(\mathbf{A}^a_t)_{k,v}$ denotes the element $(k,v)$ of the adjacency matrix $\mathbf{A}^a_t$.
Similar expressions of $d^b_t(v)$ and $g^b(v)$ on the background state can be given using $\mathbf{A}^b$ and $\sigma^{\mathbf{A}^b}$. The principle of real-time assimilation with graph data is illustrated in Fig.~\ref{fig:DA_network} where the virus propagation is simulated using the SIR model, as described in Sect.~\ref{sec: Graph-based vaccination} between two vaccination steps. Compared to the overlapped graph, the advantage of working with temporal networks is that the temporal correlation could be considered. In fact, an individual can be active for a relatively short period of time only, as shown below in Sect.~\ref{sec:Preliminary study}. Therefore, instead of using an overlapped graph (if available), analysing temporal networks can result in an efficient real-time vaccination strategy. 

A major challenge of implementing DA algorithms with graph data is the computational cost since the adjacency matrix $\mathbf{A}_t$, considered as the state variable, is a two-dimensional vector. We can rely on the assumption of graph sparsity and appropriate parameterization to reduce the computational burden. 
In this work, we propose two DA frameworks for dynamic networks updating, respectively introduced in Sect.~\ref{sec: school} and~\ref{sec: multilayer}. The former aims to reconstruct the full network with observations of sub-graphs, while the latter attempts to adjust the parameterized community-wise infectious probability, relying on multi-layer modelling. These two modellings, relatively at the local and global scale, also show the flexibility of this data assimilation framework.

\section{Numerical experiments in real-world social contact networks }
\label{sec: school}
\subsection{Assumptions and preliminary analysis}
\label{sec:Preliminary study}

This study is based on recently (before the COVID outbreak) collected face-to-face contact data from a French high school \cite{Genois2018}, which has been used to simulate a COVID outbreak \cite{Mauras2020}. The connection networks of 329 students (coverage of 86$\%$ of the students) in a high school in Lyon are available for 7374 time steps in a week. For the sake of simplicity, we condense the dynamic graph to 78 time steps by overlapping every 100 consecutive networks. Each time condensed time step 
symbolizes $30 \sim 60$ minutes. The temporal networks remain sparse since the average graph density (i.e. number of non-zero edges divided by the number of node pairs) is equal to 0.76$\%$. All contact networks are assumed to be undirected, which means the associated adjacency matrices are all symmetric (i.e., $\mathbf{A}_t = \mathbf{A}^T_t$) and the virus could spread in both directions of an edge. According to \cite{Mauras2020}, the infectious probability (of a 20-second contact) in this network can be estimated as $p \approx 0.1\% \sim 1\%$.  However, this estimated probability might be contested for the newly discovered SARS-CoV-2 variants \cite{Hou1464}. In this paper, in order to adequately investigate the optimality of different vaccination strategies, we fix the infectious probability to $p = 2\%$. The average recovery period in the SIR model is set to 60 time steps (around 4 to 5 days), following a uniform probability distribution, i.e. $T_{\gamma}\sim \textrm{unif}(55,65)$.\\

\begin{figure}[ht!]
  \centering
    \includegraphics[width = 5.5 in]{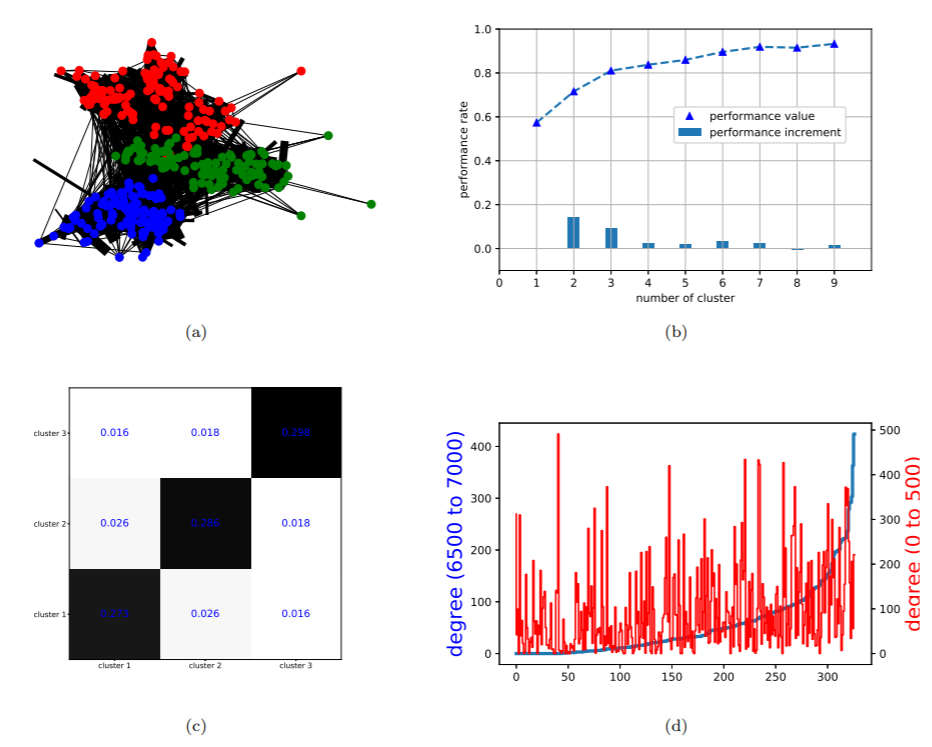}
    \caption{Preliminary analysis of the dynamical high-school connection network: (a): Overlapped contact network. (b): Performance rate $p^r (\mathcal{C})$ against assumed community number (c): Reordered adjacency matrix after clustering (d): Node degree distribution of the first and the last 50 time steps.}
  \label{fig:over_net}
\end{figure}

We begin by performing some preliminary analysis of the network data in order to better understand the underlying graph structures. The overlapped network (i.e. $\sum^{78}_{t = 1} \mathbf{A}_t$) of all the time steps is shown in Fig. \ref{fig:over_net}(a) where a clear community structure can be observed. Identifying these communities is crucial to simulating the disease spread \cite{Kiss2005}, especially for a highly infectious virus like SARS-CoV-2 \cite{Prem2020}, and to determining optimal vaccination strategies. Much effort has been given to developing community-detection algorithms in social networks \cite{agbehadji2021clustering,pares2018}. In this work, we make use of the Fluid community detection algorithm proposed by \cite{pares2018}, which is advantageous for sparse graphs since the algorithm complexity is {\em linear} to the number of non-zero edges in the network, i.e. $\mathcal{O}(|E|)$.\\

In real applications, specifying the number of communities is usually difficult. Here, we apply several times the community detection algorithms against different assumed community numbers  $k_c$, before evaluating the performance rate $p^r (\mathcal{C})$ \cite{FORTUNATO2010} of the obtained partition $\mathcal{C}$. The latter is defined as 
\begin{equation}
p^r (\mathcal{C})= \frac{|E_c|+ \big(n(n-1) -|E_{\bar{c}}| \big)}{\frac{1}{2}n(n-1)}.
\end{equation}
where $|E_c|, |E_{\bar{c}}|$ indicate the number of edges of intra- and inter-clusters respectively. The performance rate is commonly used as an indicator for finding the optimal community number, which is a standard approach for graph clustering problems. According to the result presented in Fig.~\ref{fig:over_net}(b), where we clearly observe a stationary performance rate starting from $k_c = 4$, we choose to proceed with the optimal number of clusters $k^o_c = 3$. The final clustering result is displayed in Fig.~\ref{fig:over_net}(a) where clusters/communities are shown in red, green and blue. The three detected communities are equivalently distributed, as shown by the reordered adjacency matrix (Fig.~\ref{fig:over_net}(c)), with 106, 110 and 111 nodes respectively. From a practical perspective, these communities could be considered as different grades or classes in the high school, with a similar structure to the graph data presented in \cite{Guclu2016}. \\

\subsection{DA modelling and numerical results}
\begin{figure}[H]
  \centering
    \includegraphics[width = 5.9 in]{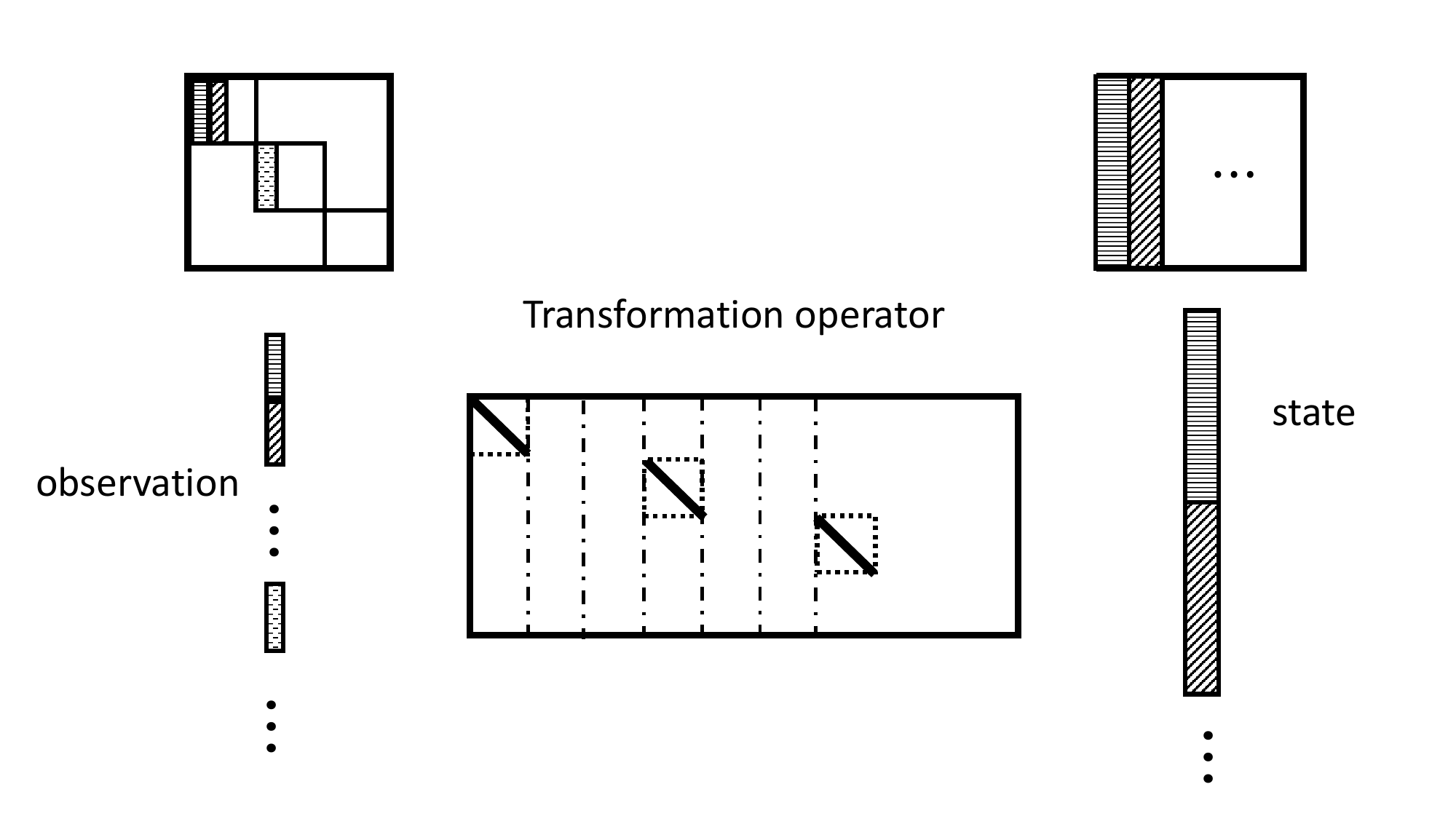}
    \caption{DA process at the current time where intra-connections are observable}
  \label{fig:infect_over}
\end{figure}

Since it is infeasible to collect contact networks via wireless equipment in all educational settings post lockdown, the objective of this study is to enhance the vaccination strategy when only partial/noisy information is available, for instance, via tracing applications. For this reason, the full contact networks $\mathbf{A}^\textrm{true}_t$ are supposed to be out of reach.
In terms of background states and observations, we suppose that the temporal network is only partially observable \textit{a priori} where 50\% to 70\%  of nodes are missing in the background estimation of the network $ \mathbf{A}^b_t \in \mathbb{R}^{329 \times 329}$. The missing nodes are selected randomly and kept invariant at all time steps. In reality, the missing nodes could refer to, for example, people who haven't installed the tracing application on their smartphone. We also use an observation vector $\mathbf{y}_t$, which contains the sub-networks for each of these three detected clusters. Thus, we suppose that the intra-community contacts of students in each class/grade are fully observable with $\mathbf{y}_t$. The objective is to perform DA algorithms sequentially to correct the knowledge of the background network relying on the observed sub-networks. The transformation operator $\textbf{H}$ is thus linear (sub-Identity matrix) and the DA problem is solved via BLUE, as shown in Eq.~\ref{eq:op_3dvar}. $\mathbf{x}^t_b  = \textrm{vect} (\mathbf{A}^b_t)$ and $\mathbf{y}_t$ are vectorized with Identity error covariances $\mathbf{B}$ and $\mathbf{O}$, as demonstrated in Fig.~\ref{fig:infect_over}. \\

After each vaccination, the SIR model is applied to simulate the virus propagation until the next time step, as summarized in Eq.~\ref{eq:SIR_graph}. An essential advantage of BLUE-type formulation with invariant prior covariances is that the Kalman gain matrix can be computed offline \textit{a priori} since it is invariant to the current $\mathbf{x}_b$ and $\mathbf{y}$. The computational cost of DA can thus be considerably reduced. The vaccination capacity is fixed $2\% (= 6 \hspace{1mm} \textrm{individuals}$ of all students for all strategies (random, highest degree, highest centrality) presented in Sect.~\ref{sec: Graph-based vaccination}, based on prior or assimilated graphs. 

\begin{figure}[ht]
  \centering
  \makebox[\linewidth][c]{
          \subfloat[]{\includegraphics[width = 2.5 in]{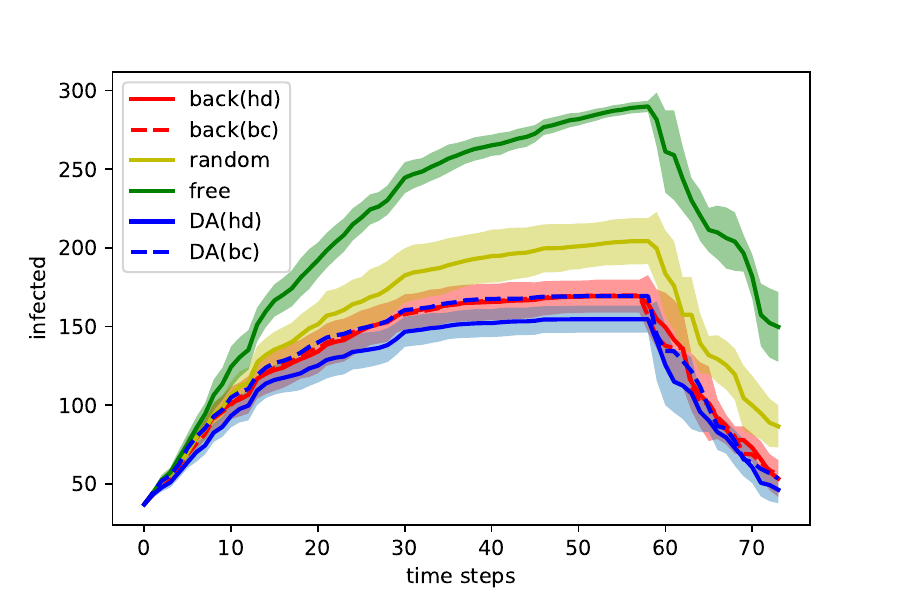}}
          \subfloat[]{\includegraphics[width = 2.5 in]{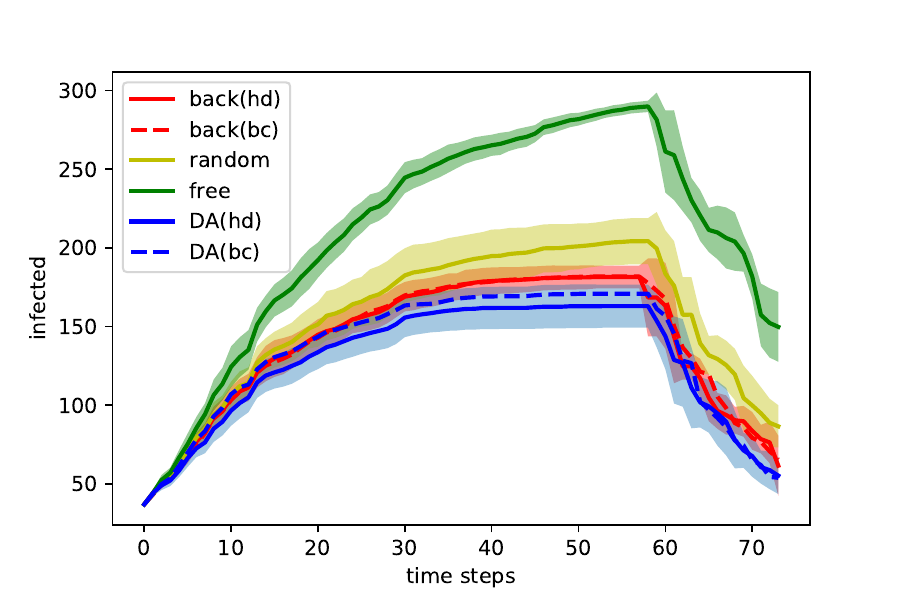}}
          \subfloat[]{\includegraphics[width = 2.5 in]{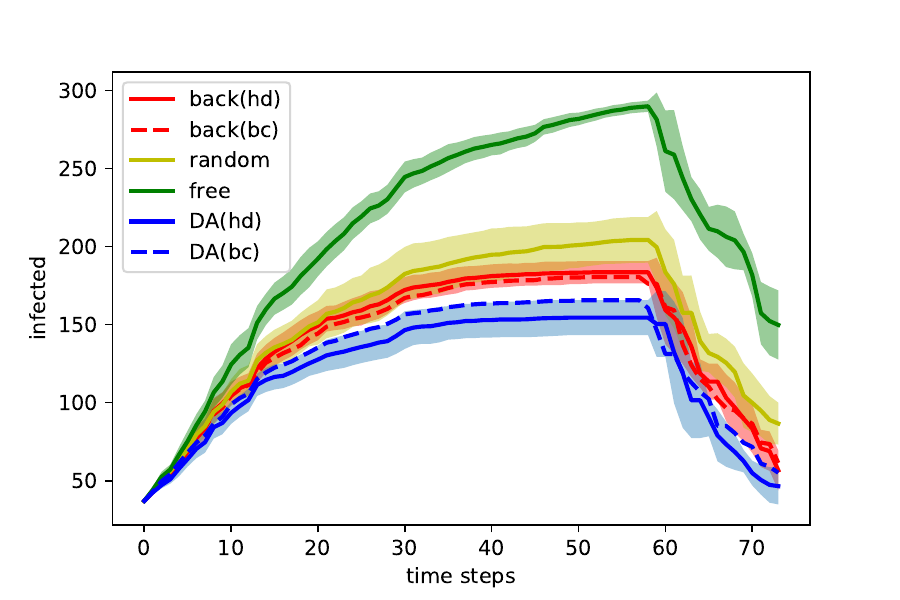}}}
    \caption{Evolution of infected against different prior error level (percentage of unobserved vertices):  (a) 50\%, (b) 60\%, (c) 70\%. Standard deviations are also displayed by transparent shades.}
  \label{fig:I_evolution}
\end{figure}

The evolution of the number of infected $|\mathbf{I}_t|$, according to different vaccination strategies, is displayed in Fig.~\ref{fig:I_evolution}, where the percentage of missing nodes in the background state is fixed as $50\%$, $60\%$ and $70\%$ respectively. To acquire robust numerical results, each type of simulation with or without vaccinations is repeated 10 times and the average values are drawn in solid or dashed curves in Fig.~\ref{fig:I_evolution}. Standard deviations of the simulations (except dashed lines) are also displayed in transparent shades to ensure the robustness of the comparison. The averaged maximum number of infected for each strategy is shown in Table~\ref{table:2}. We note that vaccinations take place at every time step for 6 selected students ($\approx 2\%$ of the population) after the simulation of virus propagations with a infectious probability of $2\%$ for each temporal edge. The initial infected $\mathbf{I}_{t=0}$, commonly used for all simulations, is randomly simulated with a probability of $P\big((\mathbf{I}_{t=0})_k)\big) = 10\%$ for $k = 1,\dots ,329$. \\

From Fig.~\ref{fig:I_evolution}, we observe that almost all averaged curves rise to a high point and peak around $t=50 - 60$ when all individuals are either infected or vaccinated. Since the vaccination process takes place in a relatively short period (a week), we suppose that the infected individuals are not detected in real-time. As a consequence, a student can be vaccinated after being infected by the virus, leading to vaccine failure. This fact emphasizes the importance of the vaccination strategy chosen. What can be clearly observed from Fig.~\ref{fig:I_evolution} is the decreasing infected number according to the vaccination strategy in the order of free (no vaccination) $\rightarrow$ random $\rightarrow$ background $\rightarrow$ assimilated (DA). This order is globally consistent regardless of time. 
First, all vaccination strategies manage to significantly reduce the number of infected and delay virus propagation compared to the free simulation (green curve). In terms of maximum infected number, for all three cases, the peak value is reduced on average by $26\%, 34\%, 34\%, 40\%$ and $37\%$, respectively for random, background with highest degree, background with highest centrality, assimilated with highest degree and assimilated with highest centrality. All other strategies are dominated by the assimilated curves, especially when proceeding with the highest degree strategy. The difference, 
in particular between background and assimilated curves, is more significant when working with large-scale networks. On the other hand, for background-network-based strategies, a growth of maximum infected number against prior error level is noticed in Table~\ref{table:2} while the results based on assimilated networks remain robust. This fact promotes the use of data assimilation on network data when prior error level can not be precisely specified. We note that the missing nodes at each time step are generated independently with no temporal correlation, explaining why reasonably good results can be obtained with $70\%$ missing nodes. In summary, numerical results show that the DA-based real-time updating of networks considerably improves the impact of vaccination, resulting in reducing virus spread. \\

In these experiments, the use of node degree (solid curves) and betweenness centrality, for both background (red) and assimilated (blue) cases, exhibits a similar performance. Such fact suggests a high-level (non-negligible) inter-clusters connections where a contrary case can be found in Sect.~\ref{sec: multilayer}. 

\begin{table}[H]
\centering
\begin{tabular}{ ||c||p{2.5cm}|p{2.5cm}|p{2.5cm}||}
 \hline
\diagbox[width=10em]{strategy}{ prior error \\ level}   & 50 \% & 60 \% & 70 \%\\
 \hline
 \hline
Free   & 88\% & 88\% & 88\% \\
 \hline
Random & 62\% & 62\% & 62\%\\
 \hline
Background (hd) & 51\% & 55\% & 56\% \\
 \hline
Background (bc)  & 52\% & 55\% & 55\% \\
 \hline
Assimilated (hd)  & 47\% & 50\% & 47\% \\
 \hline
Assimilated (bc)  & 51\%& 52\%& 50\% \\
  \hline
 \hline
\end{tabular}
\caption{Maximum number of infected (in percentage) against different vaccination strategies}
\label{table:2}
\end{table}

\section{Experiments with multi-layer networks}
\label{sec: multilayer}

\subsection{Multi-layer modelling of scale-free networks}
\begin{figure}[ht]
  \centering
   \makebox[\linewidth][c]{
        \subfloat[]{\includegraphics[width = 2.8 in]{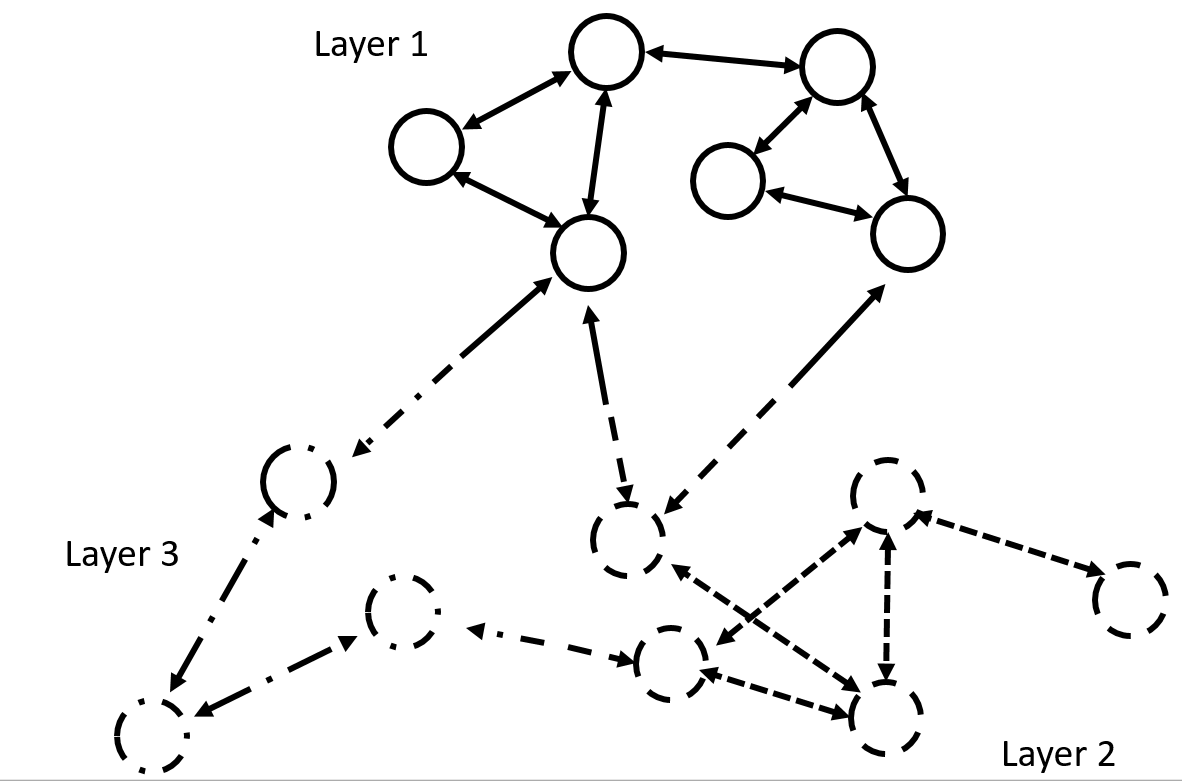}}
        \hspace{2.5cm}
        \subfloat[]{\includegraphics[width = 2.7 in]{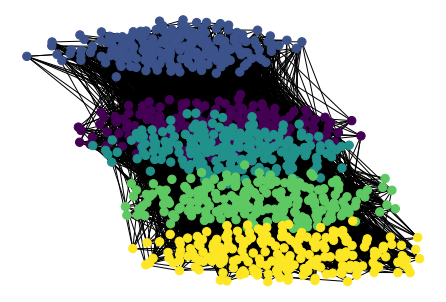}}}
    \caption{ (a): Illustration of multi-layers network modelling where the infectious probability depends on the layer of the reception node. (b): Different layers in one temporal contact network where, for example, the yellow layer could represent the community of academic staff in the department of Computing at Imperial College London and the other layers stand for students of different grades ($\textbf{CI}_a$).}
  \label{fig:infect_layer}
\end{figure}

As stated in recent research \cite{Levin2020}, the infectious probability of COVID-19 can differ significantly for different populations, based on their age, gender, activities and so on. For example, both the transmissibility and the mortality rate is reported to be higher for aged people, necessitating appropriate strategies to protect this fraction of the population. SARS-CoV-2 variants may also vary geographically \cite{Baric2020}, leading to inhomogeneous transition probabilities. Since the outbreak of the COVID-19 pandemic, continuous effort has been made to understand the behaviour of the virus infection with respect to individual-level (e.g. aged people \cite{Mueller2020}) and community-level (e.g. healthcare workers \cite{Shaukat2020}) characteristics.
These phenomena have led to the idea of using multi-layer networks, where different types of connections exist between graph nodes (see Fig.~\ref{fig:infect_layer}(a)) to simulate the virus spread in social networks. In general, multi-layer (also known as "multiplex") networks \cite{DeDomenico2016} are widely used to study graph diffusion problems \cite{Gueuningmartin2018} and define generalized versions of Pagerank \cite{DeDomenico2015}. Recently, multi-layer modelling has also been applied to COVID-19 spread simulation \cite{SCABINI2021125498} where each layer refers to a potential contamination community, such as school, workplace or transport. Appropriate use of the information on these layers can optimise vaccination strategies as mentioned in \cite{Buckner2020}, by prioritising the populations with high risk and high transmissibility.\\

Since the collection of large-scale face-to-face contact multi-layer dynamic networks is extremely complicated, we rely on conceptual modelling in this work to further examine the performance of the novel approach. Dynamic contact networks of 1000 individuals and 5 layers (each of 200 nodes) are synthetically generated, where each layer suggests a specific group in the population, according to their age or activities (e.g. students, healthcare workers). Assuming all the edges in the temporal networks are fully observable, our objective is to calibrate the time-variant infection probabilities $\{ p_{i,t} \}_{i = 1,\dots ,5}$ based on the observation of infected number in each of the layers $\{ \mathbf{I}_{i,t} \}_{i = 1,\dots ,5}$. The temporal variance of $\{ p_{i,t} \}_{i = 1,\dots ,5}$ can be a consequence of SARS-CoV-2 mutations. More precisely, the values of $\{ p_{i,t} \}_{i = 1,\dots ,5}$ update every 5 time steps, following a stochastic process,

\begin{align}
    p_{i,5t_m + 1} =  \textrm{max} (p_{i,5t_m} + \delta_{p,m}, 0) \quad \textrm{for} \quad t_m \in \mathbb{N}
\end{align}

where $\delta_{p,m} \sim unif(-0.04\%, 0.04\%)$ and the observation vector consists of incremental infected numbers $ \Delta I_{i,t} = I_{i,t} - I_{i,t-1} $. For inter-layer connections, the infectious probability is determined by the layer of the receiving nodes, i.e.

\begin{align}
    \mathcal{IP}(\mathcal{G}_t)_{i,j} = \mathcal{IP}\big((\mathbf{A}_t)_{i,j} \times p_{i,t}\big ),
\end{align}

as shown in Fig.~\ref{fig:infect_layer}(a). It is worth mentioning that the associated adjacency matrix $\mathbf{A}_t$ is no longer symmetric under this assumption. Nevertheless, the network virus spread modelling in Sect.~\ref{sec: Graph-based vaccination} remains valid.\\

As for the generation of temporal networks, we depend on the concept of scale-free networks \cite{newman2010networks} where the degree distribution follows a power law,
\begin{align}
    P_\textrm{sf}(k) \sim k ^{-\gamma} \label{eq:scale_free}
\end{align}
 where $P_\textrm{sf}(k)$ stands for the probability of a node to have $k$ connections while $2 \leq \gamma \leq3$ is a chosen parameter. To simulate intra-connections in each layer,  we use the Barabasi-Albert (BA) model \cite{Albert2002}, which is scale-free with $ \gamma = 3$, incorporating two important concepts in graph theory: growth and preferential attachment \cite{Krapivsky2008}, which exist widely in social networks. Therefore, the BA model is a reference tool to generate real-world-like networks, including web connections or citation networks. To generate a BA network, nodes are added to the network consecutively where the probability of the new node to be connected with the existing node $v$ writes
 
 \begin{align}
     P_{BA}(v) = \frac{d(v)}{\sum_j d(j)}. \label{eq:BA}
 \end{align}
The denominator in Eq.~\ref{eq:BA} represents twice the current number of edges in the network. Individuals with a higher degree have a stronger ability to grab links added to the BA network, which is an adequate assumption for social networks. Moreover, the inter-layer connections are generated randomly with a density of $0.5\%$, much sparser than intra-layer edges. Eventually, an example of a complete temporal network is drawn in Fig.~\ref{fig:infect_layer}(b) where the five layers are shown in different colors. \\

Since temporal edges are supposed to be known in this modelling, we aim to estimate $\{ p_{i,t}\}_{i = 1..5}$ based on the evolution of the infected number in all five layers. In fact, we can predict $\Delta \{I_{i,t}\}_{i = 1..5}$ via a prior estimation of $\{ p_{i,t} \}$, establishing a state-observation mapping $\mathbf{H} \in \mathbb{R}^{5 \times 5}$ for DA algorithms. The DA problem could be addressed as
\begin{align}
\textbf{x}^b = \begin{pmatrix}
        p^b_{1,t} \\
        p^b_{2,t} \\
        p^b_{3,t} \\
        p^b_{4,t} \\
        p^b_{5,t} 
        \end{pmatrix}, \quad
\textbf{y} = \begin{pmatrix}
        \Delta I_{1,t} \\
        \Delta I_{2,t} \\
        \Delta I_{3,t} \\
        \Delta I_{4,t} \\
        \Delta I_{5,t} 
        \end{pmatrix}, \quad
\textbf{H} = 
       200 \times \textbf{N} \hspace{1mm}  (\textbf{A}_t \mathbf{I}_t) \odot (\mathbf{1}_n - \mathbf{L}_{t})
\end{align}

where 
\begin{align}
    \textbf{N} = \begin{pmatrix}
        1_{1 \times 200}, 0_{1 \times 200}, 0_{1 \times 200}, 0_{1 \times 200}, 0_{1 \times 200} \\
        0_{1 \times 200}, 1_{1 \times 200}, 0_{1 \times 200}, 0_{1 \times 200}, 0_{1 \times 200} \\
        0_{1 \times 200}, 0_{1 \times 200}, 1_{1 \times 200}, 0_{1 \times 200}, 0_{1 \times 200}  \\
        0_{1 \times 200}, 0_{1 \times 200}, 0_{1 \times 200}, 1_{1 \times 200}, 0_{1 \times 200}  \\
        0_{1 \times 200}, 0_{1 \times 200}, 0_{1 \times 200}, 0_{1 \times 200}, 1_{1 \times 200}  
        \end{pmatrix}.
\end{align}

The simulation/vaccination framework is similar to the one in Sect.~\ref{sec: school} with a vaccination rate of $\approx 2\%$ of the population at each time step. This means that all people will be vaccinated before $t =50$. For all assimilations, the error covariances are set to be identity matrices, as in Sect.~\ref{sec: school}. Our goal is to determine an optimal vaccination order based on available noisy information. In order to cover more possible scenarios, we set various initial probabilities $\{ p_{i,0} \}$, as shown in Table~\ref{table:5}, denoted as $\textbf{CI}_a, \dots ,\textbf{CI}_f$. For the sake of simplicity, $\{ p_{i,0}\}$ always follow a decreasing order from layer 1 to layer 5. Typically, the initial probabilities in $\textbf{CI}_f$ are more homogeneous compared to $\textbf{CI}_a$ or $\textbf{CI}_e$. To give an example,  $\textbf{CI}_a$ could be used to simulate, for instance, a scenario in the department of computing at Imperial College where nearly 800 students plus faculty members can be found. The layer with high infectious probability may consist of professors, (senior) researchers and HR officers, while the other four layers can represent graduate or undergraduate students of different grades. The former community has a much higher average age, in contrast to the latter. Furthermore, each community holds a dense intra-connections, coherent with our model assumption. The diversity of the initial conditions ($\textbf{CI}_a, \dots ,\textbf{CI}_f$) ensures the robustness of the proposed approach. 

\begin{table}[H]
\centering
\begin{tabular}{ ||c||p{2cm}|p{2cm}|p{2cm}|p{2cm}|p{2cm}||}
 \hline
 \hline
  & Layer 1 & Layer 2 & Layer 3 & Layer 4 & Layer 5\\
  \hline
$\textbf{CI}_a$ & 2.5\% & 1\% & 1\% & 1\% & 1\% \\
  \hline
$\textbf{CI}_b$ & 3.5\% & 1.5\% & 1\% & 0.5\% & 0.5\% \\
  \hline
$\textbf{CI}_c$ & 2.5\% & 2.5\% & 2.5\% & 0.5\% & 0.5\% \\
  \hline
$\textbf{CI}_d$ & 4.5\% & 1.5\% & 1\% & 0.5\% & 0.5\% \\
  \hline
$\textbf{CI}_e$ & 3.5\% & 2.5\% & 1\% & 1\% & 0\% \\
 \hline
$\textbf{CI}_f$ & 2\% & 2\% & 1.5\% & 1\% & 1\% \\
 \hline
\end{tabular}
\caption{Initial infectious probability $\{ p_{i,0} \}$ in different layers}
\label{table:5}
\end{table}

\begin{figure}[ht]
  \centering
      \includegraphics[width = 6.9 in]{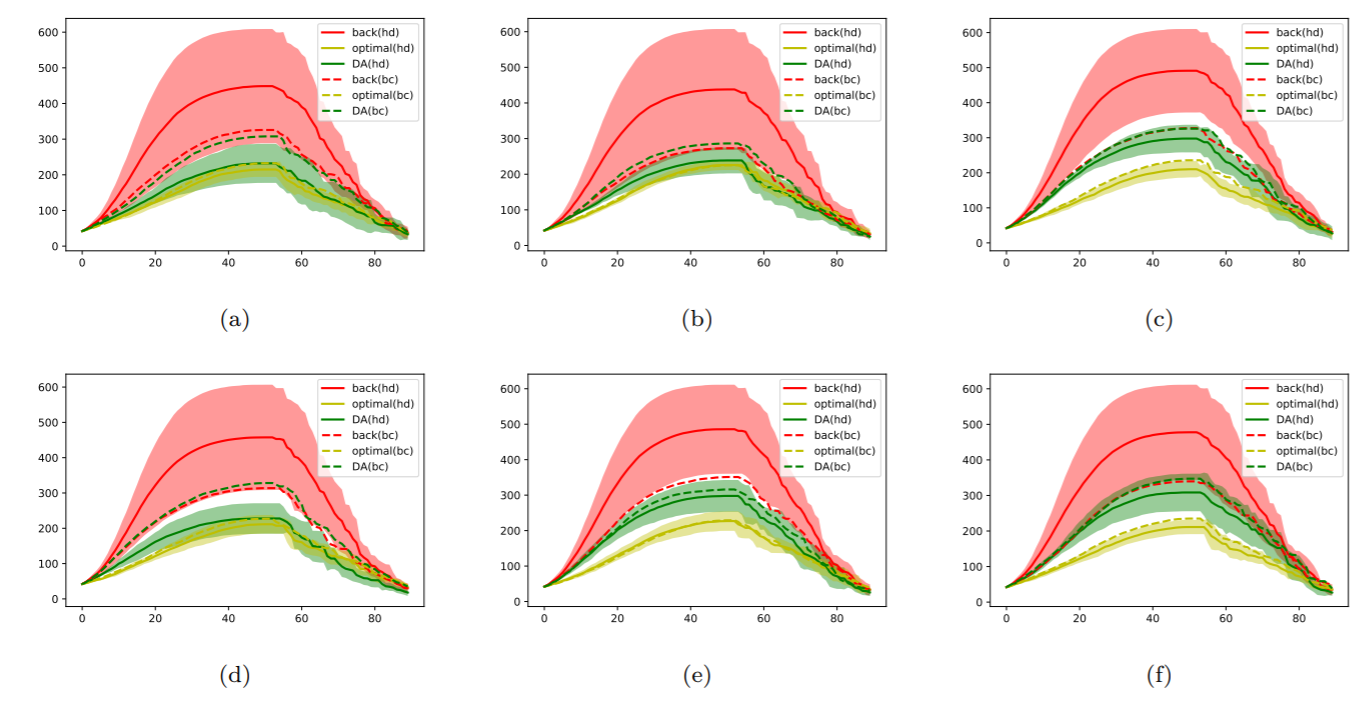}
    \caption{Evolution of infected number (average of 10 simulations) following initial conditions $\textbf{CI}_a$ .. $\textbf{CI}_f$}
  \label{fig:I_layer}
\end{figure}

The experiments set-up is similar to the one in Sect.~\ref{sec: school}. While computing the node degree and the betweenness centrality, the graph edges are weighted by either the background ($\{ p^b_{i,t} \}$) or the analyzed ($\{ p^a_{i,t} \}$) layer probabilities. Since the layer information is unattainable \textit{a priori}, background networks are set to be homogeneous (i.e.,$\{ p^b_{1,t} 	\equiv p^b_{2,t} \equiv p^b_{3,t} \equiv p^b_{4,t} \equiv p^b_{5,t}\}$). The evolution of the infected number, issued from a Monte Carlo test of 10 simulations, is illustrated in Fig.~\ref{fig:I_layer}. The stand deviation is represented by colored transparent zones. We also display the result of using exact  $\{ p_{i,t}\}$ (instead of $\{ p^b_{i,t}\}$ (red) or $\{ p^a_{i,t}\}$(green)) for vaccination in yellow. This curve is thus considered as the optimal target for the assimilation-based strategy. When vaccinating the nodes with the highest degree, a substantial advantage of the DA approach (solid green line) compared to the background one (solid red line), can be noticed in all 6 sub-figures of  Fig.~\ref{fig:I_layer}. In fact, both the maximum infected number and the average standard deviation have been significantly reduced, as confirmed in Table~\ref{table:3}. 
On the other hand, DA has much less impact when selecting the individuals with the highest centrality, as shown by the dashed lines in Fig.~\ref{fig:I_layer}. A reasonable explanation for this could be the phenomenon of brokerage \cite{Kwon2020}. The endpoints of the few inter-layer edges play an essential role in virus spread. These nodes, also known as "broker", do not necessarily have a high degree in the graph. However, since many of the shortest paths pass by them from one layer to another, the betweenness centrality may peak at these nodes with or without adjusting $\{ p_{i,t}\}$. This fact shows that when precise knowledge about inhomogeneous infectious probability is out of reach, proceeding with the highest centrality might be a robust choice. Nevertheless, both the dashed green line and the dashed red line are dominated by the solid green line (assimilated networks with the highest degree) in all 6 sub-figures.  \\

We also note that for Fig.~\ref{fig:I_layer}(a,b,d) where the five layers exhibit more variance for the initial probabilities, the assimilated curve is much closer to the optimal one. In fact, optimally vaccinating an inhomogeneous network requires less accurate knowledge of layer probabilities so long as the most infectious layers can be identified. For example, proceeding with $(5\%, 1\%, 1\%, 1\%, 1\%)$ and  $(7\%, 0.5\%, 0.5\%, 0.5\%, 0.5\%)$ for vaccine priorities may lead to similar results. \\

\begin{table}[H]
\centering
\begin{tabular}{ ||c||p{1.cm}|p{1.cm}|p{1.cm}|p{1.cm}|p{1.cm}|p{1.cm}||}
 \hline
 \hline
  & \multicolumn{6}{c}{highest degree}  \vline \\
  \hline
  & \multicolumn{3}{c}{max}  \vline & \multicolumn{3}{c}{std}  \vline \\
  \hline
    & prior & DA & true   & prior & DA & true  \\
  \hline
$\textbf{CI}_a$ & 44.9\% & 23.2\% & 21.5\% & 10.5\% & 3.7\% & 1.7\% \\
  \hline
$\textbf{CI}_b$ & 43.8\% & 23.9\% & 22.6\% & 10.9\% & 2.4\% & 1.4\%  \\
  \hline
$\textbf{CI}_c$ & 49.1\% & 29.8\% & 21.0\% & 8.1\% & 2.8\% & 1.7\% \\
  \hline
$\textbf{CI}_d$ & 45.8\% & 22.8\% & 21.1\%  & 9.7\% &  2.9\% &1.8\%\\
  \hline
$\textbf{CI}_e$ & 48.6\% & 29.8\% & 22.7\% & 8.4\% & 3.2\% & 2.0\%  \\
 \hline
$\textbf{CI}_f$ & 47.8\% & 30.9\% & 21.1\% & 9.1\% & 3.6\% & 1.6\% \\
 \hline
\end{tabular}
\caption{Averaged maximum infected number and averaged standard deviation when using node degree as order of vaccination priority}
\label{table:3}
\end{table}

\begin{table}[H]
\centering
\begin{tabular}{ ||c||p{1.cm}|p{1.cm}|p{1.cm}|p{1.cm}|p{1.cm}|p{1.cm}||}
 \hline
 \hline
  &  \multicolumn{6}{c}{highest centrality} \vline \\
  \hline
  &  \multicolumn{3}{c}{max}  \vline & \multicolumn{3}{c}{std}  \vline\\
  \hline
    & prior & DA & true   & prior & DA & true \\
  \hline
$\textbf{CI}_a$ & 32.6\% &   30.8\% &  23.3\% &  6.0\% &  5.1\% & 1.6\% \\
  \hline
$\textbf{CI}_b$ &  27.3\% & 28.6\% & 22.5\% & 2.8\% & 2.6\% & 2.2\% \\
  \hline
$\textbf{CI}_c$ &32.7\% & 32.6\% & 23.6\% & 4.4\% & 3.4\% & 2.0\% \\
  \hline
$\textbf{CI}_d$ & 31.3\% & 32.8\% &22.7\%& 30.0\% &30.0\%& 18.0\% \\
  \hline
$\textbf{CI}_e$ & 35.1\% & 31.6\% & 22.9\% & 36.3\% & 24.3\% & 17.8\% \\
 \hline
$\textbf{CI}_f$ &  34.0\% & 34.7\% & 23.5\% & 3.8\% & 3.9\% & 2.2\%\\
 \hline
\end{tabular}
\caption{Averaged maximum infected number and averaged standard deviation when using betweenness centrality as order of vaccination priority}
\label{table:4}
\end{table}

The evolution of the normalized true layer probabilities is $\frac{p_{i,t}}{\sum_k p_{k,t}}$, while their posterior (analyzed) estimation is $\frac{p^a_{i,t}}{\sum_k p^a_{k,t}}$.  The gap between the estimated and the true ratio of probabilities is rapidly reduced with the increasing of $p^a_{1,t}$, which results in a more optimal vaccination strategy. Since vaccinating infected individuals is ineffective, the early phase (around the first 20 time steps) of the outbreak is crucial to delaying the COVID spread because the most active individuals (either in terms of degree or centrality) can be infected very quickly. Therefore, the DA correction at the start of the vaccination process plays an essential role in reducing the propagation speed. On another note, we also observe that a strong oscillation in the values of $\frac{p^a_{i,t}}{\sum_k p^a_{k,t}}$ which implies high instability of the observation vector $\mathbf{y}_t = [\Delta I_{i,t}]_{i=1..5}$ due to sampling uncertainties.\\ 
In summary, the assimilation-based vaccination strategy shows competitive performance in this multi-layer modelling even though the assimilated layer probabilities are just approximations. Using the assimilated temporal networks with "highest degree" dominates other approaches, with a smaller average infected number and lower standard deviation.

\section{Conclusion and Future Work}\label{sec:conclusion}

Despite the continuous efforts, including vaccination and countrywide lockdown, it remains unclear how the COVID-19 pandemic will play out. Determining an efficient vaccination strategy is essential for combating the COVID long-term, especially with arising numbers of SARS-CoV-2 mutations. New vaccination types can be considered to fight an increasing number of SARS-CoV-2 variants. For the moment, it remains difficult to vaccinate the entire population in many countries. Using temporal contact network information can significantly improve the vaccination impact on slowing down disease propagation. This is crucial to alleviating the burden on hospitals and emergency clinics. This may also allow for the loosening of some restrictions, which is crucial to saving economies from the current pandemic. In this paper, we propose a data assimilation framework to monitor the evolution of social contact networks based on different information sources. The assimilated networks are used to govern vaccination strategies by prioritising high-risk individuals. An important strength of this framework compared to other network reconstruction methods, is the flexibility of dealing with available data and the efficiency for large-scale networks.  We have applied the proposed approach to real high school contact networks with synthetic observations and real-world-like dynamic multi-layer networks generated using the Barbasi-Albert model. The latter is used to simulate virus propagation with inhomogeneous community-level infectious probabilities. In both applications, the proposed method exhibits a significant advantage in terms of effectiveness (smaller infected number) and robustness (lower deviation). The choice of graph measures for identifying high-risk individuals, such as node degree or betweenness centrality, has also been discussed through numerical results in this study. We note that some recent work focuses on establishing data-driven models to predict individual- or community-level infection probability by learning personal data, including height, weight and health records. Computational fluid dynamics (CFD) simulations are also being developed to simulate SARS-CoV-2 transmission in schools and offices. Future work can be considered to improve individual-level modelling by incorporating these features in the contact networks. Our work opens promising perspectives on governing efficient vaccination strategies, especially for countries with a relatively low vaccination rate, or, if new vaccinations (e.g., against specific SARS-CoV-2 variants) are disseminated.
The current modelling could be extended when more network information (e.g. from tracing applications\cite{basmi2021distributed}) becomes available.

\section*{Acknowledgements}
 This work is supported by the EP/V036777/1 Risk EvaLuatIon fAst iNtelligent Tool (RELIANT) for COVID19 and the EP/T000414/1 PREdictive Modelling with QuantIfication of UncERtainty for MultiphasE Systems (PREMIERE). This research was partially funded by the Leverhulme Centre for Wildfires, Environment and Society through the Leverhulme Trust, grant number RC-2018-023.

\section*{Code and data availability}
Code for the proposed approach and the generated data is available at https://github.com/DL-WG/network\_COVID

\section*{Conflict of interest statement}
We declare that we have no financial and personal relationships with other
people or organizations that can inappropriately influence our work, there is no
professional or other personal interest of any nature or kind in any product,
service and/or company that could be construed as influencing the position
presented in, or the review of, the manuscript entitled.

\bibliographystyle{abbrv}
\bibliography{main}

\begin{thebibliography}{10}

\bibitem{agbehadji2021clustering}
I.~E. Agbehadji, R.~C. Millham, A.~Abayomi, J.~J. Jung, S.~J. Fong, and S.~O.
  Frimpong.
\newblock Clustering algorithm based on nature-inspired approach for energy
  optimization in heterogeneous wireless sensor network.
\newblock {\em Applied Soft Computing}, 104:107171, 2021.

\bibitem{Albert2002}
R.~Albert and A.-L. Barab\'asi.
\newblock Statistical mechanics of complex networks.
\newblock {\em Reviews of Modern Physics}, 74:47--97, 2002.

\bibitem{alsdurf2020covi}
H.~Alsdurf, E.~Belliveau, Y.~Bengio, T.~Deleu, P.~Gupta, D.~Ippolito, R.~Janda,
  M.~Jarvie, T.~Kolody, S.~Krastev, T.~Maharaj, R.~Obryk, D.~Pilat, V.~Pisano,
  B.~Prud'homme, M.~Qu, N.~Rahaman, I.~Rish, J.-F. Rousseau, A.~Sharma,
  B.~Struck, J.~Tang, M.~Weiss, and Y.~W. Yu.
\newblock Covid white paper.
\newblock {\em preprint arXiv:2005.08502}, 2020.

\bibitem{anderson1991discussion}
R.~M. Anderson.
\newblock Discussion: the kermack-mckendrick epidemic threshold theorem.
\newblock {\em Bulletin of mathematical biology}, 53(1-2):1, 1991.

\bibitem{Baric2020}
R.~S. Baric.
\newblock Emergence of a highly fit sars-cov-2 variant.
\newblock {\em New England Journal of Medicine}, 383(27):2684--2686, 2020.
\newblock PMID: 33326716.

\bibitem{basmi2021distributed}
W.~Basmi, A.~Boulmakoul, L.~Karim, and A.~Lbath.
\newblock Distributed and scalable platform architecture for smart cities
  complex events data collection: Covid19 pandemic use case.
\newblock {\em Journal of Ambient Intelligence and Humanized Computing},
  12(1):75--83, 2021.

\bibitem{Block2020}
P.~Block, M.~Hoffman, I.~J. Raabe, J.~B. Dowd, C.~Rahal, R.~Kashyap, and M.~C.
  Mills.
\newblock Social network-based distancing strategies to flatten the covid-19
  curve in a post-lockdown world.
\newblock {\em Nature Human Behaviour}, 4(6):588--596, 2020.

\bibitem{Buckner2020}
J.~H. Buckner, G.~Chowell, and M.~R. Springborn.
\newblock Optimal dynamic prioritization of scarce covid-19 vaccines.
\newblock {\em medRxiv : the preprint server for health sciences}, 2020.

\bibitem{camacho2020four}
D.~Camacho, {\'A}.~Panizo-LLedot, G.~Bello-Orgaz, A.~Gonzalez-Pardo, and
  E.~Cambria.
\newblock The four dimensions of social network analysis: An overview of
  research methods, applications, and software tools.
\newblock {\em Information Fusion}, 63:88--120, 2020.

\bibitem{Carrassi2017}
A.~Carrassi, M.~Bocquet, L.~Bertino, and G.~Evensen.
\newblock Data assimilation in the geosciences: An overview of methods, issues,
  and perspectives.
\newblock {\em Wiley Interdisciplinary Reviews: Climate Change}, 9(5):e535,
  2018.

\bibitem{Cauchemez2011}
S.~Cauchemez, A.~Bhattarai, T.~L. Marchbanks, R.~P. Fagan, S.~Ostroff, N.~M.
  Ferguson, and D.~Swerdlow.
\newblock Role of social networks in shaping disease transmission during a
  community outbreak of 2009 h1n1 pandemic influenza.
\newblock {\em Proceedings of the National Academy of Sciences},
  108(7):2825--2830, 2011.

\bibitem{Yiping2008}
Y.~Chen, G.~Paul, S.~Havlin, F.~Liljeros, and H.~E. Stanley.
\newblock Finding a better immunization strategy.
\newblock {\em Physical Review Letters}, 101:058701, 2008.

\bibitem{cheng2019}
S.~Cheng, J.-P. Argaud, B.~Iooss, D.~Lucor, and A.~Pon{\c{c}}ot.
\newblock Background error covariance iterative updating with invariant
  observation measures for data assimilation.
\newblock {\em Stochastic Environmental Research and Risk Assessment},
  33(11):2033--2051, 2019.

\bibitem{cheng2020b}
S.~Cheng, J.-P. Argaud, B.~Iooss, D.~Lucor, and A.~Ponçot.
\newblock Error covariance tuning in variational data assimilation: application
  to an operating hydrological model, accepted for publication in
  \textit{Stochastic Environmental Research and Risk Assessment}, 2020.

\bibitem{cheng2020graph}
S.~Cheng, J.-P. Argaud, B.~Iooss, A.~Pon{\c{c}}ot, and D.~Lucor.
\newblock A graph clustering approach to localization for adaptive covariance
  tuning in data assimilation based on state-observation mapping.
\newblock {\em arXiv preprint arXiv:2001.11860}, 2020.

\bibitem{COOPER2020}
I.~Cooper, A.~Mondal, and C.~G. Antonopoulos.
\newblock A \uppercase{SIR} model assumption for the spread of covid-19 in
  different communities.
\newblock {\em Chaos, Solitons \& Fractals}, 139:110057, 2020.

\bibitem{Davies2020}
N.~G. Davies, P.~Klepac, Y.~Liu, K.~Prem, M.~Jit, C.~A.~B. Pearson, B.~J.
  Quilty, A.~J. Kucharski, H.~Gibbs, S.~Clifford, A.~Gimma, K.~van Zandvoort,
  J.~D. Munday, C.~Diamond, W.~J. Edmunds, R.~M. G.~J. Houben, J.~Hellewell,
  T.~W. Russell, S.~Abbott, S.~Funk, N.~I. Bosse, Y.~F. Sun, S.~Flasche,
  A.~Rosello, C.~I. Jarvis, R.~M. Eggo, and C.~C.-.~w. group.
\newblock Age-dependent effects in the transmission and control of covid-19
  epidemics.
\newblock {\em Nature Medicine}, 26(8):1205--1211, 2020.

\bibitem{DeDomenico2016}
M.~De~Domenico, C.~Granell, M.~A. Porter, and A.~Arenas.
\newblock The physics of spreading processes in multilayer networks.
\newblock {\em Nature Physics}, 12(10), 2016.

\bibitem{DeDomenico2015}
M.~De~Domenico, A.~Sol{\'e}-Ribalta, E.~Omodei, S.~G{\'o}mez, and A.~Arenas.
\newblock Ranking in interconnected multilayer networks reveals versatile
  nodes.
\newblock {\em Nature Communications}, 6(1):6868, 2015.

\bibitem{sym12101646}
M.~De~la Sen, A.~Ibeas, and R.~P. Agarwal.
\newblock On confinement and quarantine concerns on an seiar epidemic model
  with simulated parameterizations for the covid-19 pandemic.
\newblock {\em Symmetry}, 12(10), 2020.

\bibitem{Durrett4491}
R.~Durrett.
\newblock Some features of the spread of epidemics and information on a random
  graph.
\newblock {\em Proceedings of the National Academy of Sciences},
  107(10):4491--4498, 2010.

\bibitem{Evensen2020}
G.~Evensen, J.~Amezcua, M.~Bocquet, A.~Carrassi, A.~Farchi, A.~Fowler, P.~L.
  Houtekamer, C.~K. Jones, R.~J. de~Moraes, M.~Pulido, C.~Sampson, and F.~C.
  Vossepoel.
\newblock An international initiative of predicting the sars-cov-2 pandemic
  using ensemble data assimilation.
\newblock {\em Foundations of Data Science}, 0(2639-8001), 2020.

\bibitem{Gracia2017}
J.~Fern{\'a}ndez-Gracia, J.-P. Onnela, M.~L. Barnett, V.~M. Egu{\'i}luz, and
  N.~A. Christakis.
\newblock Influence of a patient transfer network of us inpatient facilities on
  the incidence of nosocomial infections.
\newblock {\em Scientific Reports}, 7(1):2930, 2017.

\bibitem{Firth2020}
J.~Firth, J.~Hellewell, P.~Klepac, S.~Kissler, A.~Kucharski, and L.~Spurgin.
\newblock Using a real-world network to model localized covid-19 control
  strategies.
\newblock {\em Nature Medicine}, 26, 2020.

\bibitem{FORTUNATO2010}
S.~Fortunato.
\newblock Community detection in graphs.
\newblock {\em Physics Reports}, 486(3):75 -- 174, 2010.

\bibitem{Freeman1997}
L.~C. Freeman.
\newblock A set of measures of centrality based on betweenness.
\newblock {\em Sociometry}, 40(1):35--41, 1977.

\bibitem{Guclu2016}
H.~Guclu, J.~Read, C.~J. Vukotich, Jr, D.~D. Galloway, H.~Gao, J.~J. Rainey,
  A.~Uzicanin, S.~M. Zimmer, and D.~A.~T. Cummings.
\newblock Social contact networks and mixing among students in k-12 schools in
  pittsburgh, \uppercase{PA}.
\newblock {\em PLOS ONE}, 11:1--19, 2016.

\bibitem{Gueuningmartin2018}
M.~Gueuning, S.~Cheng, R.~Lambiotte, and J.-C. Delvenne.
\newblock Rock–paper–scissors dynamics from random walks on temporal
  multiplex networks.
\newblock {\em Journal of Complex Networks}, 8(2), 2019.

\bibitem{Genois2018}
M.~G{é}nois and A.~Barrat.
\newblock Can co-location be used as a proxy for face-to-face contacts?
\newblock {\em EPJ Data Science}, 7(1):11, 2018.

\bibitem{Genois2014}
M.~Génois, C.~Vestergaard, J.~Fournet, A.~Panisson, I.~Bonmarin, and
  A.~Barrat.
\newblock Data on face-to-face contacts in an office building suggest a
  low-cost vaccination strategy based on community linkers.
\newblock {\em Network Science}, 3:326 -- 347, 2014.

\bibitem{harling_onnela_2018}
G.~Harling and J.-P. Onnela.
\newblock Impact of degree truncation on the spread of a contagious process on
  networks.
\newblock {\em Network Science}, 6(1):34–53, 2018.

\bibitem{Hou1464}
Y.~J. Hou, S.~Chiba, P.~Halfmann, C.~Ehre, M.~Kuroda, K.~H. Dinnon, S.~R.
  Leist, A.~Sch{\"a}fer, N.~Nakajima, K.~Takahashi, R.~E. Lee, T.~M. Mascenik,
  R.~Graham, C.~E. Edwards, L.~V. Tse, K.~Okuda, A.~J. Markmann, L.~Bartelt,
  A.~de~Silva, D.~M. Margolis, R.~C. Boucher, S.~H. Randell, T.~Suzuki, L.~E.
  Gralinski, Y.~Kawaoka, and R.~S. Baric.
\newblock Sars-cov-2 d614g variant exhibits efficient replication ex vivo and
  transmission in vivo.
\newblock {\em Science}, 370(6523):1464--1468, 2020.

\bibitem{ihler2007graphical}
A.~T. Ihler, S.~Kirshner, M.~Ghil, A.~W. Robertson, and P.~Smyth.
\newblock Graphical models for statistical inference and data assimilation.
\newblock {\em Physica D: Nonlinear Phenomena}, 230(1-2):72--87, 2007.

\bibitem{Ismail2020}
S.~Ismail, V.~Saliba, J.~Bernal, M.~Ramsay, and S.~Ladhani.
\newblock Sars-cov-2 infection and transmission in educational settings: a
  prospective, cross-sectional analysis of infection clusters and outbreaks in
  england.
\newblock {\em The Lancet Infectious Diseases}, 2020.

\bibitem{Keeling2005}
M.~J. Keeling and K.~T. Eames.
\newblock {{N}etworks and epidemic models}.
\newblock {\em Journal of the Royal Society Interface}, 2(4):295--307, 2005.

\bibitem{Kiss2005}
I.~Z. Kiss, D.~M. Green, and R.~R. Kao.
\newblock {{D}isease contact tracing in random and clustered networks}.
\newblock {\em Proceedings of the Royal Society B: Biological Sciences},
  272(1570):1407--1414, 2005.

\bibitem{KOSKINEN2013514}
J.~H. Koskinen, G.~L. Robins, P.~Wang, and P.~E. Pattison.
\newblock Bayesian analysis for partially observed network data, missing ties,
  attributes and actors.
\newblock {\em Social Networks}, 35(4):514 -- 527, 2013.

\bibitem{Krapivsky2008}
P.~Krapivsky and D.~Krioukov.
\newblock Scale-free networks as preasymptotic regimes of superlinear
  preferential attachment.
\newblock {\em Physical review. E, Statistical, nonlinear, and soft matter
  physics}, 78:026114, 2008.

\bibitem{kumar2021strategy}
V.~M. Kumar, S.~R. Pandi-Perumal, I.~Trakht, and S.~P. Thyagarajan.
\newblock Strategy for covid-19 vaccination in india: the country with the
  second highest population and number of cases.
\newblock {\em npj Vaccines}, 6(1):1--7, 2021.

\bibitem{Kwon2020}
S.-W. Kwon, E.~Rondi, D.~Z. Levin, A.~D. Massis, and D.~J. Brass.
\newblock Network brokerage: An integrative review and future research agenda.
\newblock {\em Journal of Management}, 46(6):1092--1120, 2020.

\bibitem{LALMUANAWMA2020}
S.~Lalmuanawma, J.~Hussain, and L.~Chhakchhuak.
\newblock Applications of machine learning and artificial intelligence for
  covid-19 (\uppercase{SARS}-\uppercase{C}o\uppercase{V}-2) pandemic: A review.
\newblock {\em Chaos, Solitons \& Fractals}, 139:110059, 2020.

\bibitem{Levin2020}
A.~T. Levin, W.~P. Hanage, N.~Owusu-Boaitey, K.~B. Cochran, S.~P. Walsh, and
  G.~Meyerowitz-Katz.
\newblock Assessing the age specificity of infection fatality rates for
  covid-19: systematic review, meta-analysis, and public policy implications.
\newblock {\em European Journal of Epidemiology}, 35(12):1123--1138, 2020.

\bibitem{lopez2021effectiveness}
J.~Lopez~Bernal, N.~Andrews, C.~Gower, E.~Gallagher, R.~Simmons, S.~Thelwall,
  J.~Stowe, E.~Tessier, N.~Groves, G.~Dabrera, et~al.
\newblock Effectiveness of covid-19 vaccines against the b. 1.617. 2 (delta)
  variant.
\newblock {\em New England Journal of Medicine}, 2021.

\bibitem{Mauras2020}
S.~Mauras, V.~Cohen-Addad, G.~Duboc, M.~D. la~Tour, P.~Frasca, C.~Mathieu,
  L.~Opatowski, and L.~Viennot.
\newblock Analysis of mitigation of covid-19 outbreaks in workplaces and
  schools by hybrid telecommuting.
\newblock {\em medRxiv}, 2020.

\bibitem{MCCARTY2007300}
C.~McCarty, P.~D. Killworth, and J.~Rennell.
\newblock Impact of methods for reducing respondent burden on personal network
  structural measures.
\newblock {\em Social Networks}, 29(2):300 -- 315, 2007.

\bibitem{Meyers2006}
L.~Meyers.
\newblock Contact network epidemiology: Bond percolation applied to infectious
  disease prediction and control.
\newblock {\em Bulletin of the American Mathematical Society}, 44:63--86, 2006.

\bibitem{Mills2021}
M.~C. Mills and D.~Salisbury.
\newblock {{T}he challenges of distributing {C}{O}{V}{I}{D}-19 vaccinations}.
\newblock {\em EClinicalMedicine}, 31:100674, 2021.

\bibitem{Mueller2020}
A.~L. Mueller, M.~S. McNamara, and D.~A. Sinclair.
\newblock {{W}hy does {C}{O}{V}{I}{D}-19 disproportionately affect older
  people?}
\newblock {\em Aging (Albany NY)}, 12(10):9959--9981, 2020.

\bibitem{Nadler2019}
P.~Nadler, R.~Arcucci, and Y.~Guo.
\newblock Data assimilation for parameter estimation in economic modelling.
\newblock In {\em 15th International Conference on Signal-Image Technology \&
  Internet-Based Systems (SITIS) 2019}, pages 649--656, 2019.

\bibitem{Nadler2020}
P.~Nadler, S.~Wang, R.~Arcucci, X.~Yang, and Y.~Guo.
\newblock An epidemiological modelling approach for covid-19 via data
  assimilation.
\newblock {\em European Journal of Epidemiology}, 35(8):749--761, 2020.

\bibitem{Newman2002}
M.~Newman.
\newblock Spread of epidemic disease on networks.
\newblock {\em Physical review. E, Statistical, nonlinear, and soft matter
  physics}, 66:016128, 2002.

\bibitem{newman2010networks}
M.~Newman.
\newblock {\em Networks: An Introduction}.
\newblock Oxford University Press, 2010.

\bibitem{pares2018}
F.~Par{\'e}s, D.~G. Gasulla, A.~Vilalta, J.~Moreno, E.~Ayguad{\'e}, J.~Labarta,
  U.~Cort{\'e}s, and T.~Suzumura.
\newblock Fluid communities: A competitive, scalable and diverse community
  detection algorithm.
\newblock In {\em Complex Networks {\&} Their Applications VI}, pages 229--240,
  Cham, 2018. Springer International Publishing.

\bibitem{Peixoto2019}
T.~P. Peixoto.
\newblock Network reconstruction and community detection from dynamics.
\newblock {\em Physical Review Letters}, 123:128301, 2019.

\bibitem{Prem2020}
K.~Prem, Y.~Liu, T.~W. Russell, A.~J. Kucharski, R.~M. Eggo, N.~Davies, M.~Jit,
  P.~Klepac, S.~Flasche, S.~Clifford, C.~A.~B. Pearson, J.~D. Munday,
  S.~Abbott, H.~Gibbs, A.~Rosello, B.~J. Quilty, T.~Jombart, F.~Sun,
  C.~Diamond, A.~Gimma, K.~van Zandvoort, S.~Funk, C.~I. Jarvis, W.~J. Edmunds,
  N.~I. Bosse, and J.~Hellewell.
\newblock {{T}he effect of control strategies to reduce social mixing on
  outcomes of the {C}{O}{V}{I}{D}-19 epidemic in {W}uhan, {C}hina: a modelling
  study}.
\newblock {\em Lancet Public Health}, 5(5):e261--e270, 2020.

\bibitem{Rushmore2014}
J.~Rushmore, D.~Caillaud, R.~J. Hall, R.~M. Stumpf, L.~A. Meyers, and
  S.~Altizer.
\newblock {{N}etwork-based vaccination improves prospects for disease control
  in wild chimpanzees}.
\newblock {\em Journal of the Royal Society Interface}, 11(97):20140349, 2014.

\bibitem{SCABINI2021125498}
L.~F. Scabini, L.~C. Ribas, M.~B. Neiva, A.~G. Junior, A.~J. Farfán, and O.~M.
  Bruno.
\newblock Social interaction layers in complex networks for the dynamical
  epidemic modeling of covid-19 in brazil.
\newblock {\em Physica A: Statistical Mechanics and its Applications},
  564:125498, 2021.

\bibitem{Shaukat2020}
N.~Shaukat, D.~M. Ali, and J.~Razzak.
\newblock Physical and mental health impacts of covid-19 on healthcare workers:
  a scoping review.
\newblock {\em International Journal of Emergency Medicine}, 13(1):40, 2020.

\bibitem{Tillett2021}
R.~L. Tillett, J.~R. Sevinsky, P.~D. Hartley, H.~Kerwin, N.~Crawford,
  A.~Gorzalski, C.~Laverdure, S.~C. Verma, C.~C. Rossetto, D.~Jackson, M.~J.
  Farrell, S.~Van~Hooser, and M.~Pandori.
\newblock {{G}enomic evidence for reinfection with {S}{A}{R}{S}-{C}o{V}-2: a
  case study}.
\newblock {\em Lancet Infect Dis}, 21(1):52--58, 2021.

\bibitem{venkatasen2020forecasting}
M.~Venkatasen, S.~K. Mathivanan, P.~Jayagopal, P.~Mani, S.~Rajendran,
  U.~Subramaniam, A.~C. Ramalingam, V.~A. Rajasekaran, A.~Indirajithu, and
  M.~S. Somanathan.
\newblock Forecasting of the sars-cov-2 epidemic in india using sir model,
  flatten curve and herd immunity.
\newblock {\em Journal of ambient intelligence and humanized computing}, pages
  1--9, 2020.

\bibitem{Wang2020}
S.~{Wang}, X.~{Yang}, L.~{Li}, P.~{Nadler}, R.~{Arcucci}, Y.~{Huang},
  Z.~{Teng}, and Y.~{Guo}.
\newblock A bayesian updating scheme for pandemics: Estimating the infection
  dynamics of covid-19.
\newblock {\em IEEE Computational Intelligence Magazine}, 15(4):23--33, 2020.

\bibitem{wu2020comprehensive}
Z.~Wu, S.~Pan, F.~Chen, G.~Long, C.~Zhang, and S.~Y. Philip.
\newblock A comprehensive survey on graph neural networks.
\newblock {\em IEEE transactions on neural networks and learning systems},
  32(1):4--24, 2020.

\bibitem{YANG2019115}
Y.~Yang, A.~McKhann, S.~Chen, G.~Harling, and J.-P. Onnela.
\newblock Efficient vaccination strategies for epidemic control using network
  information.
\newblock {\em Epidemics}, 27:115 -- 122, 2019.

\bibitem{Yong2020}
S.~E.~F. Yong, D.~E. Anderson, W.~E. Wei, J.~Pang, W.~N. Chia, C.~W. Tan, Y.~L.
  Teoh, P.~Rajendram, M.~P. H.~S. Toh, C.~Poh, V.~T.~J. Koh, J.~Lum, N.~M.
  Suhaimi, P.~Y. Chia, M.~I. Chen, S.~Vasoo, B.~Ong, Y.~S. Leo, L.~Wang, and
  V.~J.~M. Lee.
\newblock {{C}onnecting clusters of {C}{O}{V}{I}{D}-19: an epidemiological and
  serological investigation}.
\newblock {\em Lancet Infect Dis}, 20(7):809--815, 2020.

\bibitem{you2020handling}
J.~You, X.~Ma, D.~Y. Ding, M.~Kochenderfer, and J.~Leskovec.
\newblock Handling missing data with graph representation learning.
\newblock {\em arXiv preprint arXiv:2010.16418}, 2020.

\end{thebibliography}
\end{document}